# Evidence of electron-electron interactions around Van Hove singularities of a graphene Moiré superlattice

Si-Yu Li[1,2], Ke-Qin Liu[1,2], Long-Jing Yin[1,2], Wen-Xiao Wang[1,2], Wei Yan[1,2], Xu-Qin Yang[1,2], Jun-Kai Yang[1,2], Haiwen Liu[1,2], Hua Jiang[3,*], and Lin He[1,2,*]

[1]Center for Advanced Quantum Studies, Beijing Normal University, Beijing, 100875, People's Republic of China

[2]Department of Physics, Beijing Normal University, Beijing, 100875, People's Republic of China

[3]College of Physics, Optoelectronics and Energy, Soochow University, Suzhou, 215006, People's Republic of China

Correspondence and requests for materials should be addressed to H.J. (e-mail: jianghuaphy@suda.edu.cn) and L.H. (e-mail: helin@bnu.edu.cn).

**A variety of new and interesting correlated states have been predicted in graphene monolayer doped to Van Hove singularities (VHSs) of its density-of-state (DOS). However, tuning the Fermi energy to reach a VHS of graphene by either gating or chemical doping is prohibitively difficult, owning to their large energy distance (~ 3 eV). Therefore, these correlated states, which arise from effects of strong electron-electron interactions at the VHSs, have remained experimentally elusive. Here, we report experimental evidences of electron-electron interactions around the VHSs of a twisted bilayer graphene (TBG) through scanning tunneling microscopy measurements. By introducing a small twisted angle between two adjacent graphene sheets, we are able to generate low-energy VHSs arbitrarily approaching the Fermi energy. The split of the VHSs are observed and the symmetry breaking of electronic states around the VHSs are directly visualized. These results experimentally demonstrate the important effects of electron-electron interactions on electronic properties around the VHSs of the TBG, therefore providing motivation for further theoretical and experimental studies in graphene systems with considering many-body interactions.**

The possibility that graphene could exhibit novel correlated states has attracted much attention over the years [1-7]. However, all these new phases of matter predicted in theory emerge only when the Fermi level is close to the VHSs of graphene monolayer, which, unfortunately, are too far (~ 3 eV) to be reached in experiment [8]. Therefore, an experimental study of electronic properties at the VHSs of graphene is still lacking up to now. Recently, the newly discovered twisted bilayer graphene (TBG) systems with small twisted angles [9-22] open up new prospects in this direction because their VHSs, which originate from the two low-energy saddle points in the band structures, are fully accessible with the present doping and gating techniques. Here we show that by reducing the twisted angle, it is possible to tune the VHSs of the TBG arbitrarily approaching the Fermi energy. When the VHSs is tuned close to the Fermi level, we observe the split of the VHSs and the distortion of electronic states around the VHSs. Such a result directly demonstrates the important effects of the electron-electron interactions on the electronic properties around the VHSs.

In our experiment, graphene bilayers were synthesized by chemical vapour deposition (CVD) on Rh foils (see Methods and Supplementary Fig. 1 [23]) and we have demonstrated previously that the graphene bilayers on Rh foils have a strong twisting tendency [12,20,24]. Therefore, such a system provides an ideal platform for twist engineering of the low-energy VHSs. Figure 1(a) shows a schematic structure of two misoriented graphene sheets with a twisted angle $\theta$. The period $D$ of the twist-induced moiré pattern is related to $\theta$ by $D = a/[2\sin(\theta/2)]$ with $a = 0.246$ nm the lattice constant of graphene. Figure 1(b) shows a representative scanning tunneling microscope (STM) image of a twisted bilayer with the period $D \approx 11.3$ nm (see Methods for STM measurement [23]). The twisted angle is estimated to be $\theta \approx 1.2°$. As schematically shown in Fig. 1(a), the twisting not only results in the moiré pattern, but also leads to a relative shift of the two Dirac cones $|\Delta \boldsymbol{K}| = 2|\boldsymbol{K}|\sin(\theta/2)$, where $\boldsymbol{K}$ is the reciprocal-lattice vector, in reciprocal space. A finite interlayer coupling between the two graphene sheets leads to the emergence of two saddle points appearing at the intersections of the two Dirac cones, which consequently generate two low-energy VHSs in the density-of-state (DOS), as shown in Fig. 1(c) for the theoretical local DOS (LDOS) of the TBG with twisted angle $\theta \approx 1.2°$ (see Methods and Supplementary Fig. 2 for calculation details [23]). The upper panel of Fig. 1(d) shows a typical scanning tunneling spectroscope (STS) spectrum of the TBG with the energy resolution of about 10 meV (see Methods and Supplementary Fig. 3(a) for more spectra with the similar energy resolution). The energy difference of $\Delta E_{VHS} \approx 25$ meV of the two VHSs peaks, as shown in the upper panel of Fig. 1(d), agrees well with both our theoretical result in Fig. 1 (c) and that reported in previous experimental work with a similar twisted angle [13]. However, by carefully measuring the spectra with a high spectroscopic resolution of around 1 meV (see Methods for more details about the high-resolution STS measurements), as shown in the lower panel of Fig. 1(d), we observe a completely new feature of the VHSs: each of the VHSs splits into two peaks with energy separation of about 10 meV. The splitting of the VHSs, which has never being reported before in the TBG, only depends slightly on the positions of the moiré pattern (see Supplementary Fig. 3(b) [23]). Very recently, similar splitting of DOS peaks has been observed around monovacancy and isolated hydrogen atoms on graphene and is attributed to the enhanced electron-electron interactions because of that the intensity of the localized DOS increases dramatically with approaching the monovacancy or the isolated hydrogen atom [25,26]. Therefore, the observed splitting of the VHSs indicates the lifting of degenerate electronic states at the VHSs, which may be induced by the enhanced electron-electron interaction with the VHSs approaching the Fermi level.

To further explore the origin of the splitting of the VHSs, we measure the structure of the TBG at varying voltage bias. Figure 2(a) and 2(b) show two typical STM images recorded at voltage bias away from and around the VHSs, respectively (see Supplementary Fig. 4 for more STM images recorded at different sample bias [23]). Obviously, the obtained structures of the moiré patterns in these STM images are quite different. For the STM image recorded at bias (i.e., at energy) away from the VHSs (Fig. 2(a)), the twist-induced moiré patterns are almost circular symmetry, as the schematic structure shown in Fig. 1(a). However, there is an obvious structural deformation of the moiré patterns when we measure the STM image at voltage bias (i.e., at energy) around the VHSs, as shown in Fig. 2(b). We can rule out the tip effect on the deformation of the moiré pattern [27] as the origin of the observed phenomenon because the shape of the moiré patterns does not show dependency on the varying currents in our STM measurements (see Supplementary Fig. 5 for STM images measured on various currents). To quantitatively describe the deformation, we use the full width at half maximum (FWHM) of the scanning profile lines across the moiré patterns in $x$ and $y$ directions, i.e., $L_x$ and $L_y$, to represent the shape of the moiré patterns (see

Supplementary Fig. 6(a)). Figure 2(c) and 2(d) show the measurements of the $L_x$ and $L_y$ on the moiré patterns at the energies of 100 meV and -8.5 meV, respectively (see Supplementary Fig. 6(b) and (c) for the measurements at other energies). Figure 2(f) summarizes the ratio of $L_y/L_x$ as a function of the voltage bias. Obviously, the symmetry of the moiré patterns is broken when the energy of the TBG is close to the low-energy VHSs. In the STM images, both topography and electronic states of the sample contribute to the contrast. To explore the origin of the deformation of the moiré patterns, we carry out atomic resolution STM measurements of the moiré patterns, as shown in Fig. 2(e) (also see Supplementary Fig. 7 for details). These measurements demonstrate that the lattice distortion in the moiré patterns is negligible, indicating that the observed distortion in the moiré patterns is mainly contributed by the electronic effects at the energies around the VHSs. The distortion of the electronic states, which are strongly related to the electron-electron interactions magnified by the enhanced DOS around the VHSs [1-6,28], could be directly measured by STS maps and will be discussed subsequently.

Here we should point out that the small twisted angle is vital in the enhanced electron-electron interactions around the VHSs in the TBG because of the following two main aspects. First, the VHSs are approaching the charge neutrality point of the system and also the Fermi level in our experiment with decreasing the twisted angles (see Fig. 1d and the inset of Fig. 2(e)). Second, the magnitude of the DOS at the VHSs increases and the FWHM of the VHSs decreases with decreasing the twisted angles (the inset of Fig. 2(e)). Therefore, the electron-electron interactions are expected to be more significant for the TBG with small twisted angles. To verify such a result, controlled experiments on several TBG with different twisted angles are carried out. In the TBG with larger twisted angles (the VHSs are more than 100 meV away from the Fermi energy), we do not observe the deformation of the moiré patterns at the energy of VHSs and the corresponding splitting of the VHSs in the same STM measurements (see Fig. 2(e) and Supplementary Figs. 8-10 for the results of other TBG samples with different twisted angles). The above controlled experiments also help us to remove possible effects of the STM tip and the supporting substrate as the origin of the observed phenomena.

Differential conductance maps (STS maps), which are widely used to directly reflect the spatial distribution of the LDOS [11,14,29,30], are further measured to analyze the electronic states of the TBG in our experiment. The presence of the moiré pattern in the TBG is expected to lead to spatial variation of the LDOS. Figure 3(a) and 3(b) show representative STS maps recorded at energies away from the VHSs of the TBG with $\theta \approx 1.2°$. The distribution of the LDOS (Fig. 3(a) and 3(b)) reveals the same period and circular symmetry of the moiré pattern but exhibits the inverted contrast comparing with that shown in the STM image (Fig. 1(b)). Such a result is reproduced well in our theoretical simulations, as shown in Fig. 3(c) and 3(d), reflecting the presence of the moiré potential in the twist-induced moiré pattern [21,22,32].

Around the VHSs, the distribution of the LDOS predicted by theoretical calculations (without taking into account the electron-electron interactions) also shows the same period and circular symmetry of the moiré pattern, as shown in Fig. 4(a) and 4(b). The theoretical spatial-distribution of the LDOS around the VHSs exhibits the same contrast as the moiré pattern in the STM image, which is the slight difference comparing with that shown in Fig. 3. Similar evolution of the LDOS distribution over the moiré pattern is observed in the TBG with slightly large twisted angles, for example in the TBG with $\theta \approx 1.9°$ ($D$ = 7.3 nm) in our experiment (see Supplementary Fig. 11 [23]). However, in the TBG with $\theta \approx 1.2°$, the enhanced electron-electron interactions around the low-energy VHSs are expected to affect the spatial distribution of the LDOS at VHSs. Such an effect can be directly visualized through the STS maps in our STM measurements. Figure 4(c)-4(f) show STS maps recorded at the four peaks of the VHSs in the TBG with $\theta \approx 1.2°$, exhibiting pronounced spatial modulation of the LDOS. Quite different from the LDOS distribution recorded at energies away from the VHSs (Fig.3), both the period and the symmetry of the moiré pattern are completely removed from the spatial distribution of the LDOS at the VHSs. The distortion of electronic states observed in Fig. 4(c)-(f) cannot be induced by the tip effect on STS maps [31] because that similar effect has not been observed in other TBG measured in the same experimental conditions (see Supplementary Fig. 11 for further discussion [23]). Here, we attribute the observed distorted electronic states to the charge-density-wave (CDW) [33-35] triggered by the electron-electron interactions around the VHSs of the TBG. Because of the modulation by the CDW, the spatial distribution of LDOS around the VHSs no longer follows the periodicity of the structure (here, the moiré pattern), which is in good agreement with previous studies in other CDW systems [33-35]. The distortion of the periodicity and the symmetry of the LDOS around the VHSs then affects the contrast in the STM images, resulting in the deformation of the moiré patterns recorded around the VHSs. Additionally, the observed CDWs for the occupied and empty VHSs are quite different (Fig. 4(c)-4(f)), which is also consistent with previous observations for occupied and empty states in other correlated systems with strong electron-electron interactions [33-35].

In the TBG with small twisted angles, there is pronounced Fermi velocity renormalization [36,37] and the Fermi velocity will reduce to about zero when $\theta \approx 1.1°$. Therefore, the quasiparticles are strongly localized in the TBG with $\theta \approx 1.2°$. This is also very important for the emergence of the correlated states because of that the ratio of the interaction energy compared to the kinetic energy is expected to increase dramatically around the VHSs of this TBG. By applying a magnetic field up to 13 T, we still cannot observe Landau quantization (Supplementary Fig. 10a [23]) since that the picture of two-dimensional free electron gas no longer applies in the TBG with $\theta \approx 1.2°$. Such a result consists with the previous observations for the slightly TBG in STM measurements [14,36,37]. The strong magnetic fields affect slightly the tunneling spectra (Supplementary Fig. 12 [23]) and the spatial distribution of the LDOS in the TBG (Supplementary Fig. 13 [23]). Such a result is reasonable and easy to be understood since that there may be a competition between twist-induced localization and cyclotron motion generated by the large perpendicular magnetic fields. The magnetic length $l_B = \sqrt{\hbar/eB}$ generated by the magnetic field of 13 T is estimated to be about 7 nm, which is much smaller than the period of the Moiré patterns, ~ 11.3 nm, in the TBG with $\theta \approx 1.2°$.

In summary, the observation of the splitting of the VHSs and the distorted spatial distribution of the LDOS around the VHSs demonstrate the important effects of electron-electron interactions at the VHSs in the graphene moiré superlattice. Our result provides a new route to realize novel correlated states in graphene systems in the near future.

Note added: After submission of this paper, we became aware of a related manuscript [38] showing evidences of electron-electron interactions in slightly twisted bilayer graphene through transport measurements.

Unraveling the Intrinsic and Robust Nature of van Hove Singularities in Twisted Bilayer Graphene by Scanning Tunneling Microscopy and Theoretical Analysis. *Phys. Rev. Lett*. **109**, 196802 (2012).

14. Yin, L.-J., Qiao, J.-B., Zuo, W.-J., Li, W.-T., He, L. Experimental evidence for non-Abelian gauge potentials in twisted graphene bilayers. *Phys. Rev. B* **92**, 081406(R) (2015).

15. Suarez Morell, E., Correa, J. D., Vargas, P., Pacheco, M., Barticevic, Z. Flat bands in slightly twisted bilayer graphene: tight-binding calculations. *Phys. Rev. B* **82**, 121407(R) (2010).

16. Ohta, T., Robinson, J. T., Feibelamn, P. J., Bostwick, A., Rotenberg, E., Beechem, T. E. Evidence for Interlayer Coupling and Moiré Periodic Potentials in Twisted Bilayer Graphene. *Phys. Rev. Lett*. **109**, 186807 (2012).

17. He, W.-Y., Chu, Z.-D. & He, L. Chiral Tunneling in a Twisted Graphene Bilayer. *Phys. Rev. Lett.* **111**, 066803 (2013).

18. Bistritzer, R. & MacDonald, A. H. Moire bands in twisted double-layer graphene. *Proc Natl Acad Sci* (*USA*) **108**, 12233-12237 (2011).

19. San-Jose, P., Gonzalez, J., Guinea, F. Non-Abelian gauge potentials in graphene bilayers. *Phys. Rev. Lett.* **108**, 216802 (2012).

20. Yan, W., Meng, L., Liu, M., Qiao, J.-B., Chu, Z.-D., Dou, R.-F., Liu, Z., Nie, J.-C., Naugle, D. G., He, L. Angle-dependent van Hove singularities and their breakdown in twisted graphene bilayers. *Phys. Rev. B* **90**, 115402 (2014).

21. Chu, Z.-D., He, W.-Y., He, L. Coexistence of van Hove singularities and superlattice

**Acknowledgments**

This work was supported by the National Natural Science Foundation of China (Grant Nos. 11674029, 11422430, 11374035, 11374219, 11504008), the National Basic Research Program of China (Grants Nos. 2014CB920903, 2013CBA01603, 2014CB920901), the NSF of Jiangsu province, China (Grant No. BK20160007), the program for New Century Excellent Talents in University of the Ministry of Education of China (Grant No. NCET-13-0054). L.H. also acknowledges support from the National Program for Support of Top-notch Young Professionals and support from "the Fundamental Research Funds for the Central Universities".

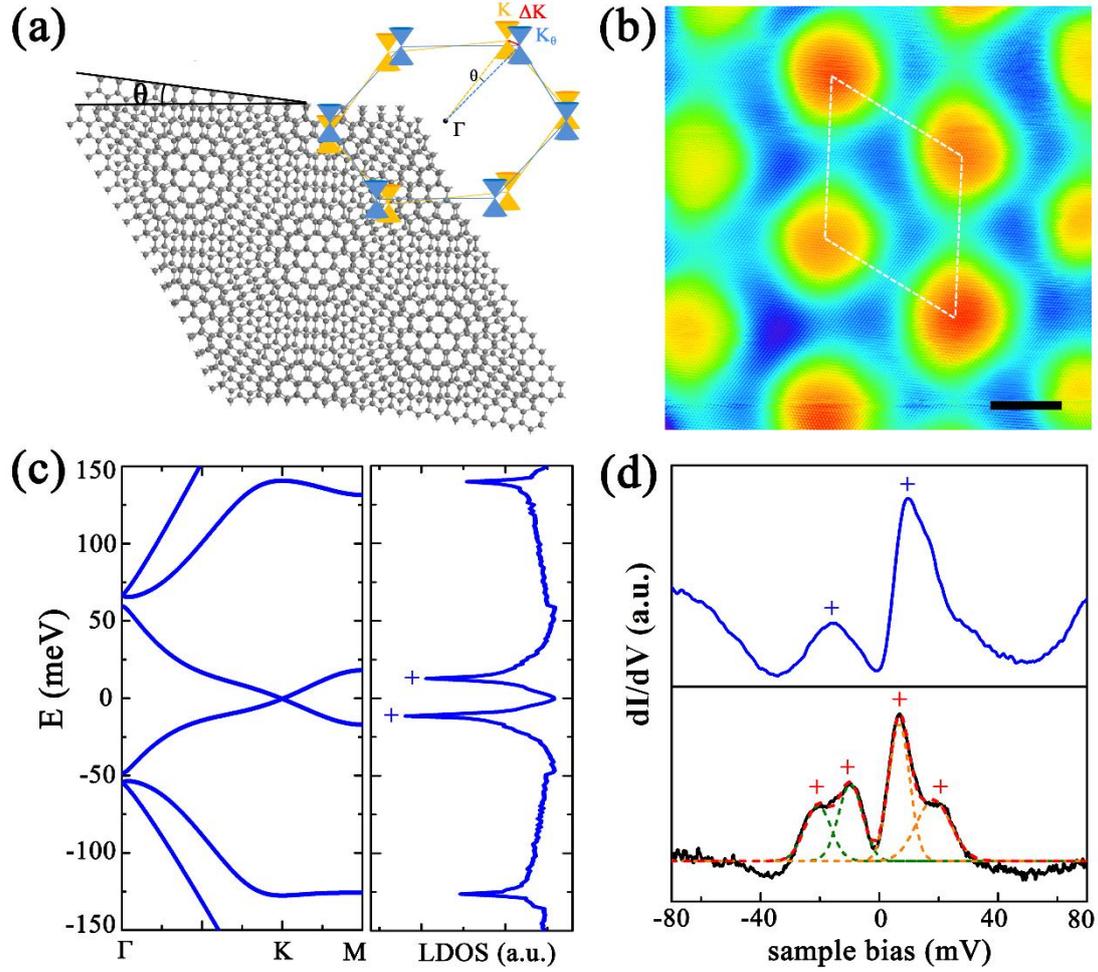

**Figure 1.** The splitting of VHSs in the TBG with twisted angle $\theta \sim 1.2°$. **(a).** Schematic structure of two misoriented graphene sheets with a twisted angle $\theta$ and schematic two Dirac cones, $K$ and $K_\theta$, of the two layers intersect in momentum space. **(b).** A 30 nm × 30 nm STM image of the TBG with the period $D \sim 11.3$ nm and the twisted angle $\theta \sim 1.2°$ ($V_{sample}$ = -78 mV and $I$ = 0.4 nA). Scale bar: 5 nm. **(c).** Theoretical calculated electronic structure of the TBG with twisted angle $\theta \sim 1.2°$ and the corresponding LDOS with two VHS peaks. **(d).** Top panel: a STS spectrum taken from the TBG in **b** with two low-energy VHS peaks. Bottom panel: a typical high-resolution STS spectrum of the same TBG with each of the two VHSs splitting into two peaks. The dashed lines mark the result of Gaussian fitting for the split VHSs.

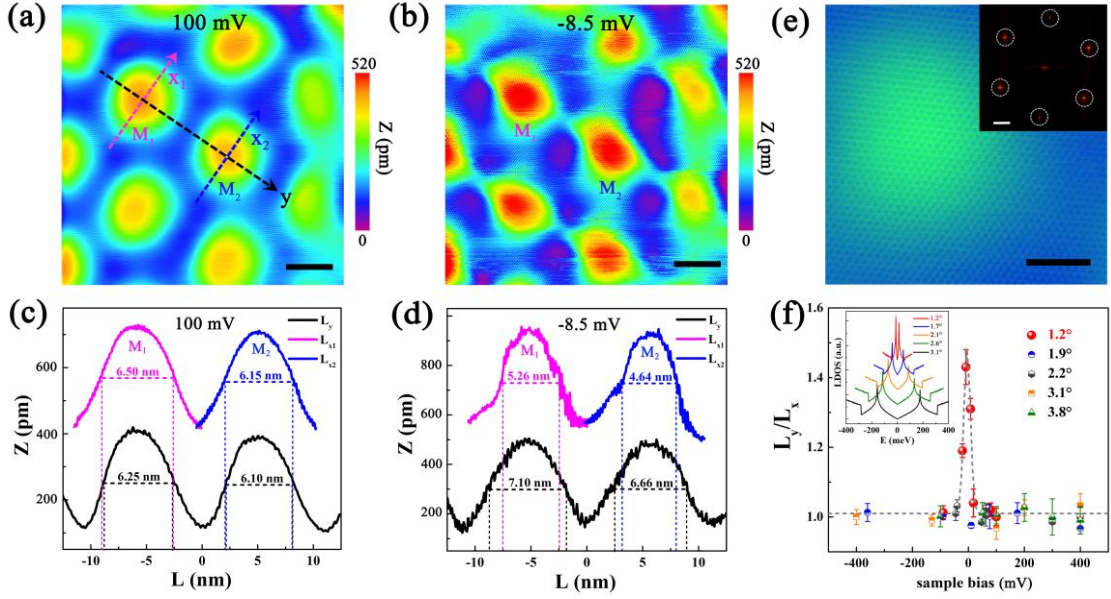

**Figure 2.** Deformation of the moiré superlattice around VHSs. **(a), (b).** The 30 nm × 30 nm STM images recorded at voltage bias 100 mV and -8.5 mV which are away from and around the VHSs respectively ($I$ = 0.2 nA). Scale bar: 5 nm. **(c), (d).** The scanning profile lines across the moiré patterns $M_1$ and $M_2$ in $x$ and $y$ direction at the energies of 100 meV and -8.5 meV, respectively. The measured directions $y$, $x_1$ and $x_2$ are marked in (a). And the numbers correspond to the measured FWHM values of the profile lines. For clarity, we stack the profile lines by y offsets. **(e).** The atomic-resolution STM image of the moire pattern ($V_{sample}$ = 300 mV and $I$ = 0.3 nA). Scale bar: 1 nm. Inset show the corresponding 2D Fourier transform image. The six bright spots marked by the white dashed circles correspond to the reciprocal lattices of graphene. **(f).** Summarization for the ratio of $L_y/L_x$ as a function of the voltage bias for several TBG samples with different twisted angles, where $L_x$ and $L_y$ are the FWHM of the moiré patterns along $x$ and $y$ directions respectively. The red dashed line is the Gaussian fitting between the $L_y/L_x$ of TBG with twisted angle $\theta \sim 1.2°$ and the voltage bias. Inset: theoretical calculated LDOS as the function of energy $E$ in the TBG with different

twisted angles.

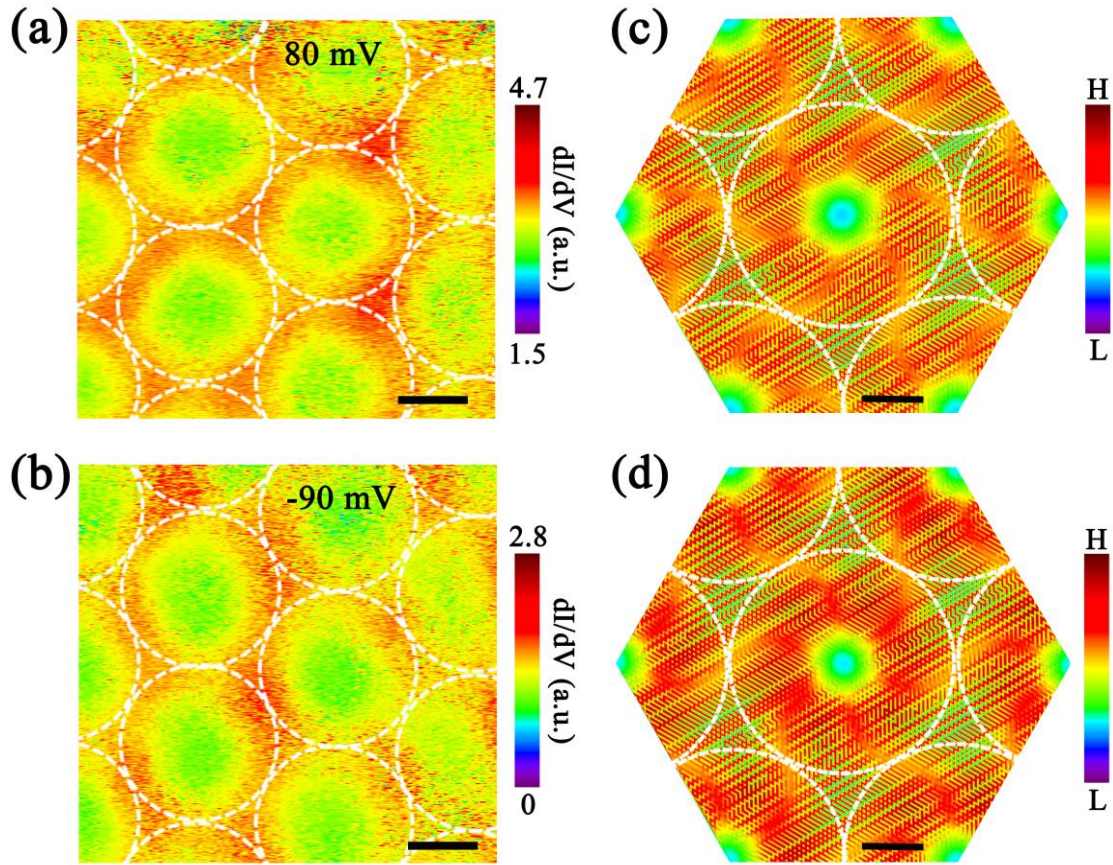

**Figure 3.** Spatial distribution of the LDOS away from the VHSs in the TBG. **(a), (b).** Experimental STS maps recorded in the TBG with twisted angle $\theta \sim 1.2°$ at the fixed sample bias of 80 mV in a and -90 mV in b, respectively. Scale bar: 5 nm. The white dashed circles mark the period and circular symmetry of the moiré patterns in STM image. **(c), (d).** Theoretically spatial distribution of the LDOS at the energies away from the VHSs, i.e., 139 meV in c and -126 meV in d, in the TBG with twisted angle $\theta \sim 1.2°$. Scale bar: 3 nm.

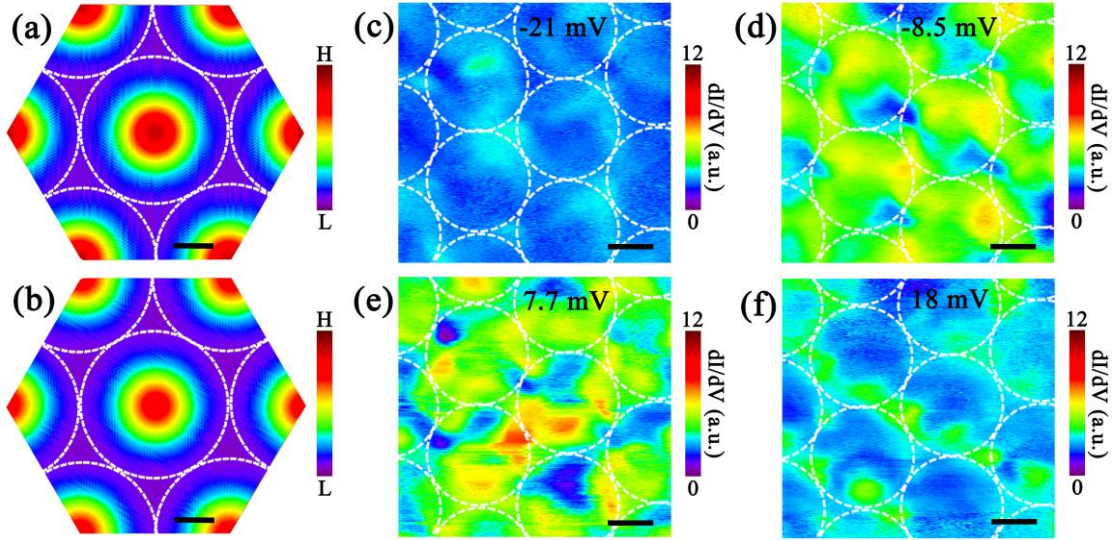

**Figure 4.** Distortion of the electronic states around the VHSs of the TBG. **(a), (b).** Theoretical spatial distribution of the LDOS in the TBG with twisted angle $\theta \sim 1.2°$ at the energies of the two VHSs in the calculated STS spectrum of Fig. 1c, respectively. Scale bar: 3 nm. **(c)-(f).** Experimental STS maps recorded in the same region of the TBG at the fixed sample bias -21 mV, -8.5 mV, 7.7 mV and 18 mV, respectively, which correspond to the four split VHSs in Fig. 1d. Scale bar: 5 nm. The white dashed circles mark the period and circular symmetry of the moiré superlattice in the STM image.